\documentclass[runningheads]{llncs}
\usepackage{graphicx}
\usepackage{amsmath}
\usepackage{siunitx}

\usepackage[inline]{fixme}

\begin{document}
\title{Automated Detection of Cortical Lesions in Multiple Sclerosis Patients with 7T MRI}
\titlerunning{Automated cortical lesion detection at 7T}

\author{Francesco La Rosa\inst{1,2,3,4} 
\and Erin S Beck\inst{4} 
\and Ahmed Abdulkadir \inst{5,6} 
\and Jean-Philippe Thiran \inst{1,3} 
\and Daniel S Reich \inst{4} 
\and Pascal Sati \inst{4,7} 
\and Meritxell Bach Cuadra \inst{3,2,1} 
}

\authorrunning{F. La Rosa et al.}

\institute{LTS5, Ecole Polytechnique F\'{e}d\'{e}rale de Lausanne, Switzerland \and  Medical Image Analysis Laboratory, CIBM, University of Lausanne, Switzerland \and Radiology Department, Lausanne University Hospital, Switzerland \and Translational Neuroradiology Section, National Institute of Neurological Disorders and Stroke, National Institutes of Health, Bethesda, MD, USA \and University Hospital of Old Age Psychiatry and Psychotherapy, University of Bern, Bern, Switzerland  \and Center for Biomedical Image Computing and Analytics, Department of Radiology, Perelman School of Medicine, University of Pennsylvania, Philadelphia, PA, USA
\and Department of Neurology, Cedars-Sinai Medical Center, Los Angeles, CA, USA }

\maketitle              

\begin{abstract}
The automated detection of cortical lesions (CLs) in patients with multiple sclerosis (MS) is a challenging task that, despite its clinical relevance, has received very little attention. Accurate detection of the small and scarce lesions requires specialized sequences and high or ultra-high field MRI. For supervised training based on multimodal structural MRI at 7T, two experts generated ground truth segmentation masks of 60 patients with 2014 CLs. We implemented a simplified 3D U-Net with three resolution levels (3D U-Net\textsuperscript{-}). By increasing the complexity of the task (adding brain tissue segmentation), while randomly dropping input channels during training, we improved the performance compared to the baseline. Considering a minimum lesion size of 0.75 \si{\micro\liter}, we achieved a lesion-wise cortical lesion detection rate of 67\% and a false positive rate of 42\%. However, 393 (24\%) of the lesions reported as false positives were {\sl post-hoc} confirmed as potential or definite lesions by an expert. This indicates the potential of the proposed method to support experts in the tedious process of CL manual segmentation.

\keywords{MRI \and  Ultra-high field \and Multiple Sclerosis \and Cortical lesions \and Segmentation \and CNN.}
\end{abstract}
\section{Introduction}
Multiple sclerosis (MS) is the most common demyelinating disease affecting the central nervous system. Demyelination results in focal lesions that appear in both the white matter (WM) and gray matter (GM) of the brain and of the spinal cord. Magnetic resonance imaging (MRI) is the conventional imaging tool used for the diagnosis and evaluation of disease progression and therapy response, with a focus on WM lesions (WMLs) dissemination in space and time. While WMLs remain a hallmark of MS, in 2017 cortical lesions (CLs) were added to the diagnostic criteria of MS \cite{criteria2017}. CLs are associated with worse disability and with progressive forms of MS, and these associations appear to be at least partially independent of WML burden \cite{calabrese,treaba}. CLs can be divided into three subtypes: leukocortical (type I, involving the cortex and WM), intracortical (type II, entirely within the cortex but not touching the pial surface), and subpial (type III and IV, touching the pial surface of the cortex). There is increasing evidence that subpial lesions form due to inflammation in the overlying meninges, in a somewhat distinct mechanism from that of WM and leukocortical lesion formation \cite{magliozzi}. Thus, it is important to understand the clinical implication of subpial lesions and their response to existing and novel MS treatments in order to optimize MS diagnostic and prognostic accuracy and maximize treatment efficacy.

CLs, however, are only visible with specialized advanced MRI sequences at high (3T) and ultra-high magnetic field (7T) \cite{kober,kilsdonk}. Specifically, 7T MRI has become the reference {\sl in vivo} technique for CL identification due to its increased signal-to-noise ratio (SNR) and often higher imaging resolution \cite{kilsdonk}. Moreover, 7T MRI is significantly more sensitive to intracortical and subpial lesions compared to 3T \cite{maranzano}. Magnetization-prepared 2 rapid acquisition with gradient echo (MP2RAGE) \cite{kober,mp2rage}, in particular, has emerged as a promising sequence for detecting CLs at 7T \cite{erin}, but different T2*-weighted (T2*w) contrasts have been suggested as well \cite{maranzano,sati}. Overall, the combination of MP2RAGE and one (or more) T2*w sequence gives the highest sensitivity \cite{erin}.

A wide range of machine learning algorithms have been proposed in order to automatically segment WMLs in MRI~\cite{review}. Recently, deep learning algorithms have achieved the best performance in terms of WML detection and segmentation \cite{challenge}. However, the automated detection of CLs in MRI has been barely explored. CLs are smaller and show a lower intensity contrast compared to WMLs. Moreover, CLs locations (within the cortex and at its interface with WM) make their detection more challenging than lesions entirely within the WM. Finally, the need for advanced MRI sequences limits the dataset availability and thus the availability of training samples. Nevertheless, at 3T, four methods have been tested to automatically segment both WMLs and CLs \cite{mj,mj_pv,larosa2019,larosa2020}. All methods were applied in a multimodal setting, including advanced sequences such as MP2RAGE, 3D fluid-attenuated inversion recovery (FLAIR) and 3D double-inversion recovery (DIR) \cite{mj,mj_pv}, or only MP2RAGE and 3D FLAIR \cite{larosa2019,larosa2020}.

At 7T, however, the problem of automatically detecting CLs differs from 3T for three main reasons. Firstly, a significantly higher number of CLs is visible, including subpial ones which are almost not seen at lower magnetic fields. Secondly, the increased inhomogeneity in the radiofrequency (B1) field and local variations of the static magnetic field (B0) create artefacts and spatial distortions. Thirdly, the T2w 3D FLAIR sequence, commonly used at 1.5 and 3T, presents artefacts and alternative advanced sequences are acquired instead. Beyond these challenges and in  the recent approval of the first 7T MRI scanner for clinical use, the development of automated tools for MS lesion segmentation at ultra-high field is needed to help physicians in better analysing those images. Pioneering work of Fartaria et al. \cite{fartaria2019} proposed an automated segmentation of MS lesions at 7T, based on the concatenation of skull stripping, tissue segmentation, and morphological operations. Their approach, based on a single MP2RAGE scan, segments both WMLs and CLs (mostly leukocortical), reporting an accuracy for CLs of 58\% with 40\% of false positives.
\begin{figure}[ht!]
\centering
\includegraphics[width=0.70\textwidth]{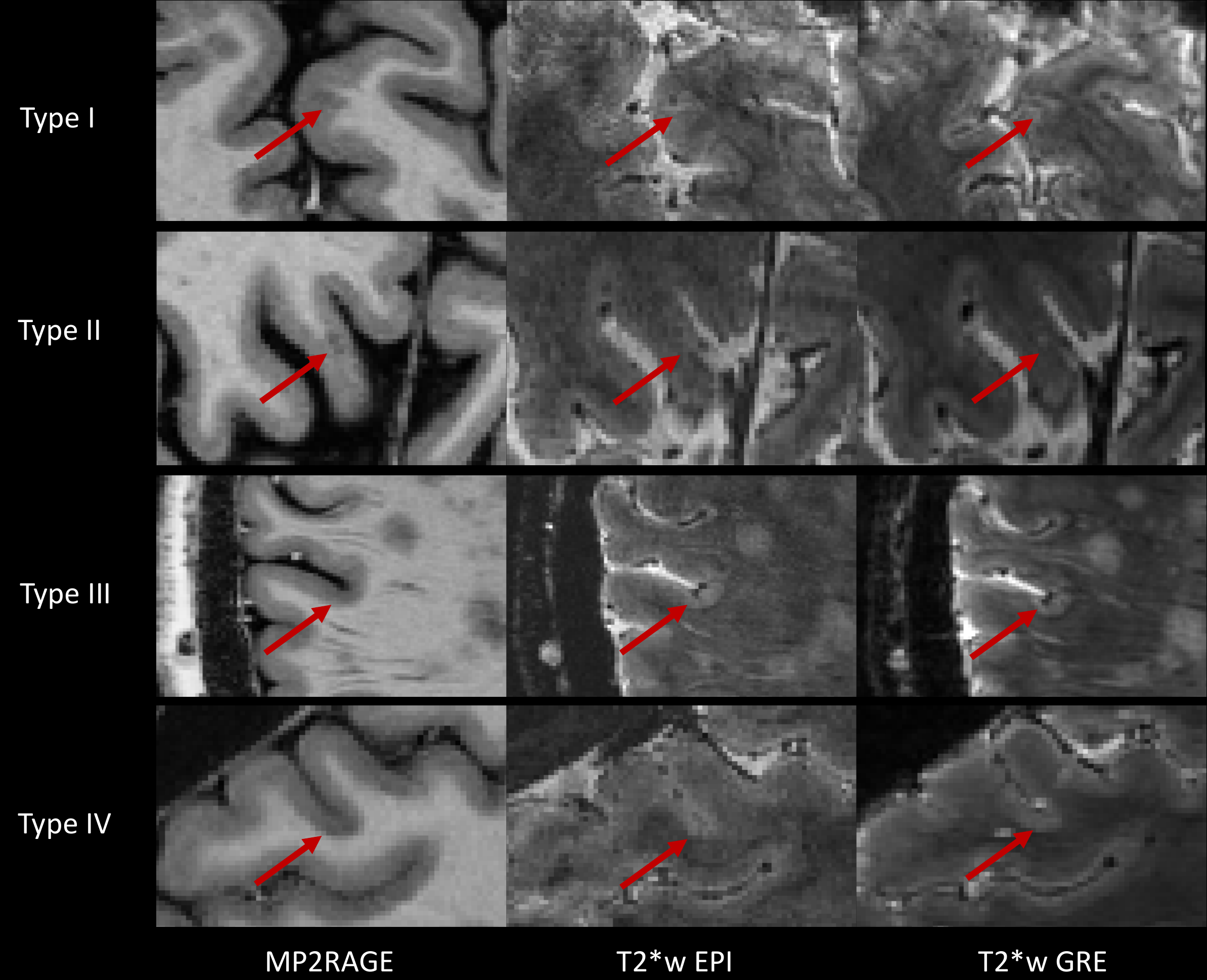}
\caption{Examples of the four types of CLs seen in the three different contrasts at 7T MRI.} \label{fig1}
\end{figure}

In our work, we present a deep learning 7T multimodal approach for CL detection only, considering MP2RAGE, T2*w echo planar imaging (EPI) \cite{sati,sati2}, and T2*w gradient recalled echo (GRE) sequences. This is the first attempt to detect CLs and classify them in two different types: leukocortical and subpial/intracortical. We implemented a 3D U-Net \cite{3dunet} with fewer resolution levels resulting in substantially fewer parameters. Considering this architecture as baseline, we increased the complexity of the prediction task by simultaneously predicting main tissue types and randomly dropping input channels during training. Another important contribution of this work is the validation framework. Compared to the only previous work in literature, we increase the number of patients and CLs used for validation: from 25 and 364 in \cite{fartaria2019} to 60 and 2014. Furthermore, we target small CLs with a minimum volume of 0.75 \si{\micro\liter} instead of 6 \si{\micro\liter} considered in \cite{fartaria2019}.

\section{Methodology}
\label{datasetSection}
\subsection{Dataset}
MRI acquisitions were done on 60 patients (38/22 female/male, $49 \pm 11$ years old, age range $[29-77]$ years) with Expanded Disability Status Scale (EDSS) scores ranging from 0 to 7.5 (median $2.0 \pm 2.0$), 17 were progressive and 43 relapsing remitting MS. Imaging was performed on a 7T whole-body research system (Siemens Healthcare, Erlangen, Germany) using a 32-channel head coil. The MRI protocol included: (i) 3D MP2RAGE \cite{kober} (TR/TI1/TI2/TE = 6000/800/2700/5 ms, voxel size = $0.5\times0.5\times0.5 mm\textsuperscript{3}$), (ii) 3D-Segmented T2*w EPI \cite{sati,sati2} (TR/TE = 52/23 ms, voxel size = $0.5\times0.5\times0.5 mm\textsuperscript{3}$) acquired in two partially overlapping volumes, (iii) T2*w multi-echo GRE (TR/TE1/TE2/TE3/TE4/TE5 = 4095/11/23/34/45/56 ms, voxel size = $0.5\times0.5\times0.5 mm\textsuperscript{3}$) acquired in three volumes.

The study was approved by the Institutional Review Board of our institution, and all patients gave written informed consent prior to participation.

\textbf{Manual CL detection and tissue segmentation.} 2014 CLs were manually detected and classified by consensus by one neurologist and one neuroradiologist, both with several years of experience identifying CLs. They analyzed multiple planes and considered, if needed, all three MRI contrasts. Additional lesions which did not fully convince the experts, due to their poor intensity contrast, small size, or appearance on a single contrast, were marked as ``possible CLs". The brain tissue was segmented in white and gray matter with the automatic software ANTs \cite{ants} for the training tissue labels.

\textbf{Types of MS CLs.} Our experts classified the CLs according to \cite{calabrese}. See Figure \ref{fig1} for an example of each type. Within our dataset, 38\% of the CLs identified belong to type I, 7\% to type II, 44\% to type III, and 11\% to type IV.

\textbf{Pre-processing. }The images of each subject were linearly registered to the same space (MP2RAGE), and intensity non-uniformities were corrected using a variant of the nonparametric nonuniform intensity normalization algorithm \cite{n4}.

\subsection{Network's details}
\begin{figure}
\centering
\includegraphics[width=0.95\textwidth]{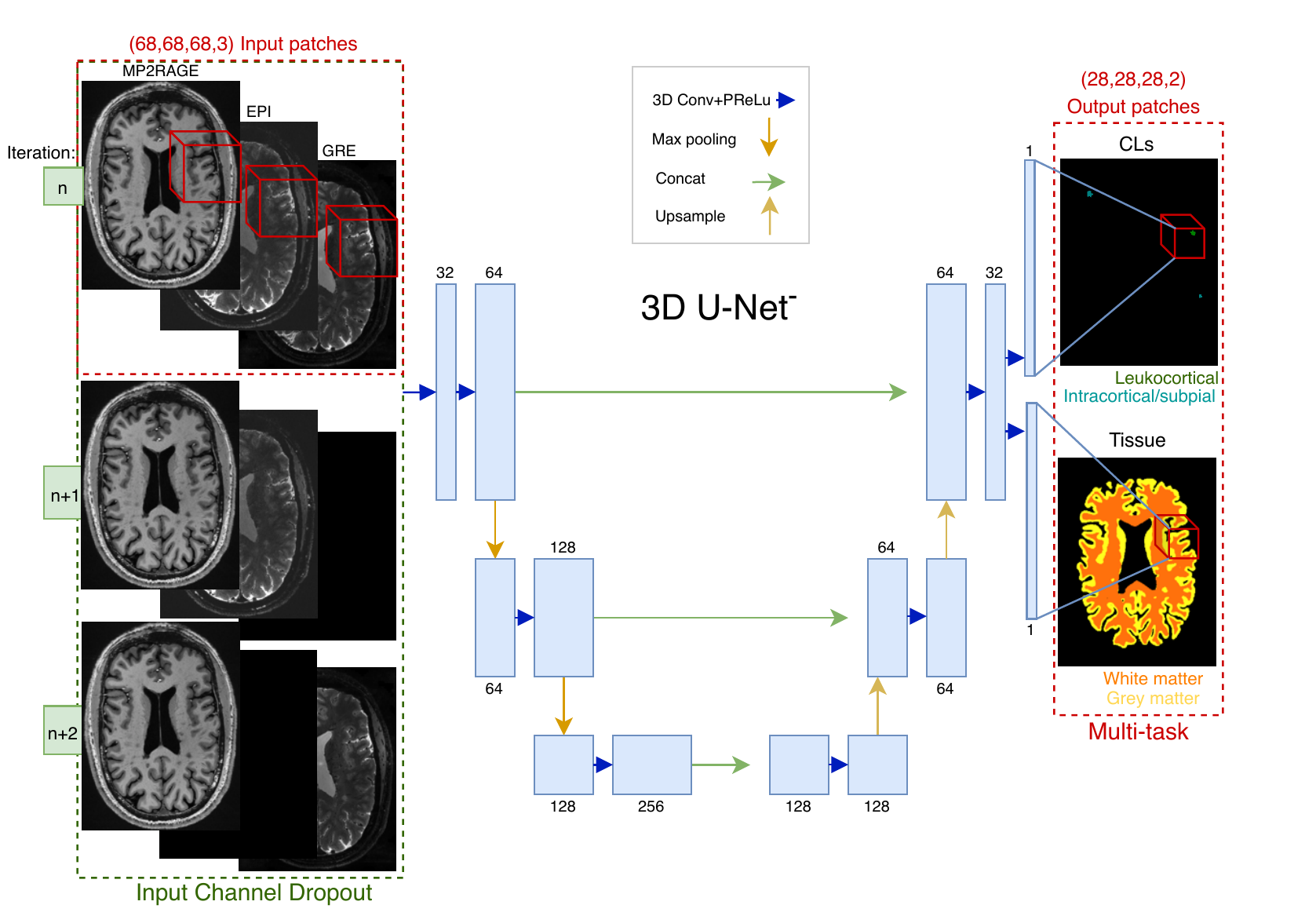}
\caption{Scheme of the 3D U-Net\textsuperscript{-} implemented. In input, the three MRI contrasts, with input channel dropout, in output a CLs mask and a tissue segmentation.} \label{fig2}
\end{figure}

The chosen network architecture (Figure \ref{fig2}) is inspired by the 3D U-Net \cite{3dunet}, which has proved successful for several biomedical imaging segmentation tasks, including MS lesion segmentation \cite{kaur}. Given the limited amount of data, to avoid overfitting we reduced the complexity of the network by removing one resolution layer, and thus the term U-Net\textsuperscript{-}. We were interested in a voxel-wise segmentation of CLs with three levels (leukocortical lesions, subpial/intracortical lesions, and background), distinguishing CLs between these primarily in the GM (type II, III, and IV) and these affecting the interface between GM and WM (type I). We chose to work with patches of 68x68x68 as input yielding an output of 28x28x28 voxels.
We propose an additional output layer (denoting this network as multi-task U-Net\textsuperscript{-}) in order to guide the learning procedure. This additional task consisted of brain tissue segmentation in WM, GM, and background. A joint tissue and lesion segmentation has proven already to be promising in \cite{reuben} regarding WM hyperintensities segmentation. Even though tissue segmentation is not the goal of this work, this architecture allows the network to be aware of the tissue location of the CLs and improved the detection metrics (see Section \ref{sec:results}). Each lesion was sampled with the same probability, regardless of its size and was balanced with samples from the rest of the brain.
The networks were trained with voxel-wise weighted cross-entropy in order to balance the two output maps. In the CLs output map weights of 15, 1, and 0 were assigned to CLs voxels, background, and WMLs respectively in order not to penalize the network if these are segmented (as some leukocortical lesions appear very similar to juxtacortical WMLs). In the tissue map, all the lesions had weight 0 to simplify the tissue segmentation in these regions, whereas other voxels had weight 1. The initial learning rate was set to 1e-4, and Adam was used as optimizer.

\subsubsection{Data augmentation}
\begin{figure}
\centering
\includegraphics[width=0.99\textwidth]{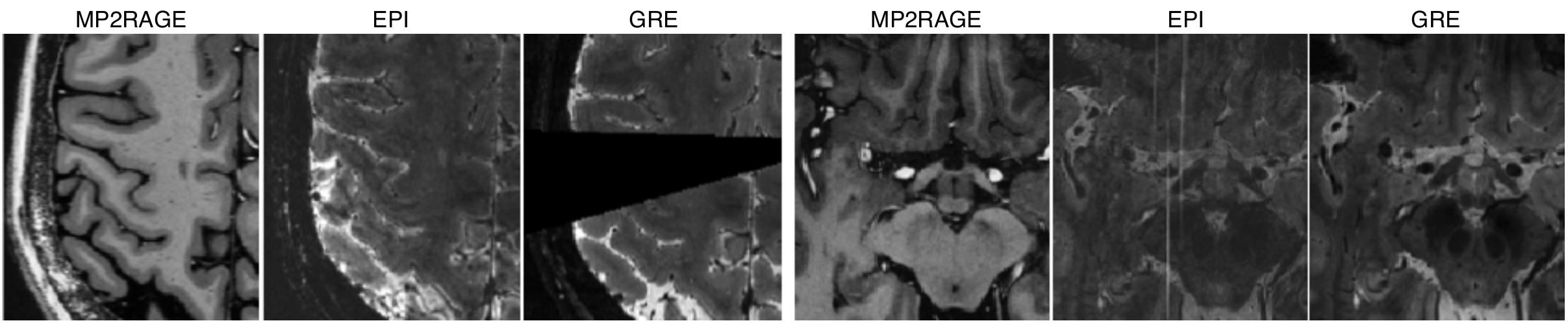}
\caption{On the left, a zoomed-in example where part of the brain is missing in the GRE sequence. On the right, artefacts affect the EPI.  } \label{fig3}
\end{figure}

We applied extensive data augmentation to tackle the risk of overfitting. Random rotations of up to 180 degrees in all three planes and flipping of the axes were applied. Input-channel dropout (ICD) for the two T2*w contrasts was also evaluated. This consists in randomly dropping (eg. multiplying it by zero) one of the two T2*w contrasts at each training iteration. The main motivation is that small portions of the brain were occasionally missing on data from the T2*w GRE sequence, whereas the T2*w EPI shows several artefacts (see Figure \ref{fig3}). We hypothesized this augmentation technique would improve the network robustness to these images' artefacts (see section \ref{sec:results}). 
\newline

Training was performed on a NVIDIA TITAN X GPU for 50000 iterations and took approximately 22 hours per each fold. The code has been implemented in NiftyNet \cite{niftynet} running on top of Tensorflow. The code and models can be obtained from our research website \footnote[1]{https://github.com/Medical-Image-Analysis-Laboratory}.

\subsection{Evaluation}
We evaluated U-Net\textsuperscript{-}, multi-task U-Net\textsuperscript{-} (adding the tissue segmentation output layer), and multi-task U-Net\textsuperscript{-} with ICD in a 6-folds cross-validation over the 60 cases available. 
We considered a minimum lesion size of 6 voxels (0.75 \si{\micro\liter}), much lower than the one previously considered by any method performing automatic MS lesion segmentation. As proposed in \cite{challenge}, we compute the following metrics: absolute volume difference (AVD), CLs lesion-wise true and false positives rates (LTPR and LFPR, respectively), CLs patient-wise true and false positives rates (TPR and FPR, respectively) and CL classification accuracy (Accuracy).
Wilcoxon signed‐rank test is performed to compare the TPR and FPR patient-wise of the different architectures. Differences are considered significant for p-value $<$ 0.05.
\section{Results}
\label{sec:results}
\begin{figure}[hbt!]
\centering
\includegraphics[width=0.99\textwidth]{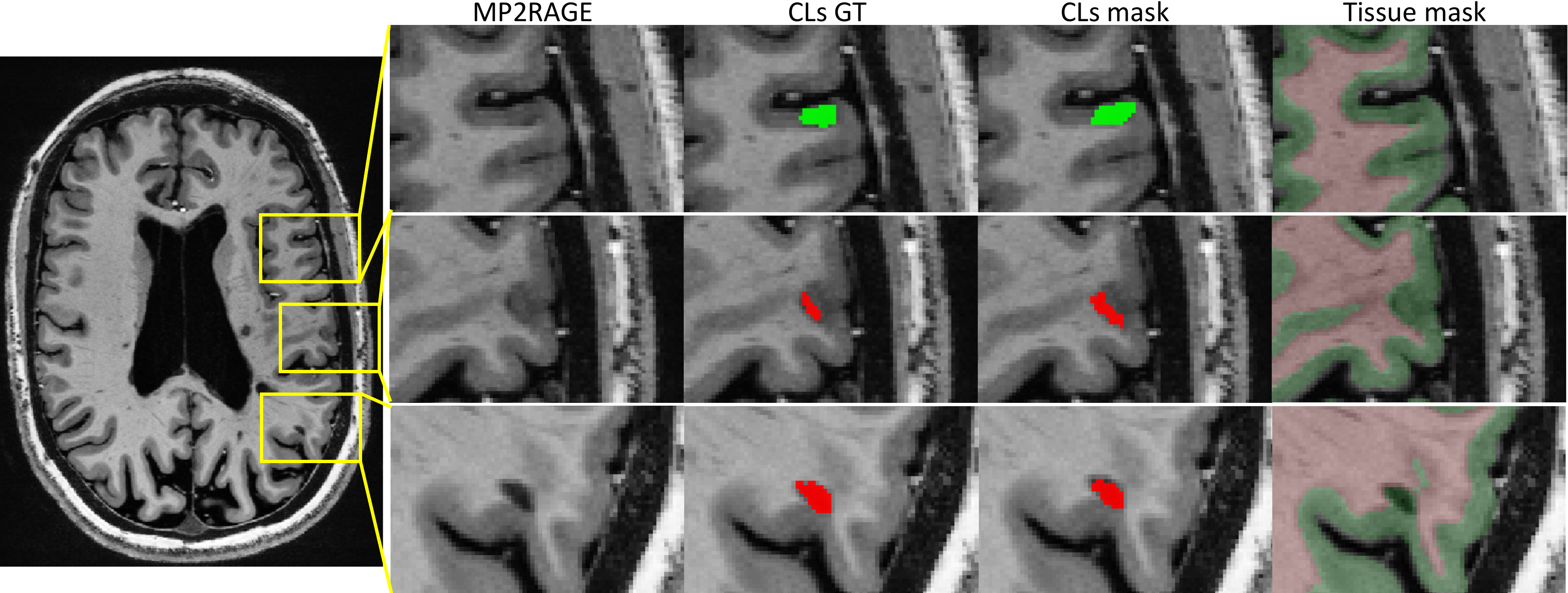}
\caption{Examples of CLs (leukocortical in red and subpial/intracortical in green) in the experts' ground truth compared to the output masks of Multi-task 3D U-Net\textsuperscript{-} + ICD. The tissue classes are color coded as red
(WM) and green (GM). \fxnote[inline]{The space here could be used more efficiently. For instance, include these inputs and outputs (in this size) into Figure 2. Figure 2 could then become the central figure illustrating inputs, weight maps, hidden feature maps, false positives that turned out to be TPs, augmentation.}} \label{fig6}
\end{figure}

\begin{figure} [hbt!]
\centering
\includegraphics[width=0.999\textwidth]{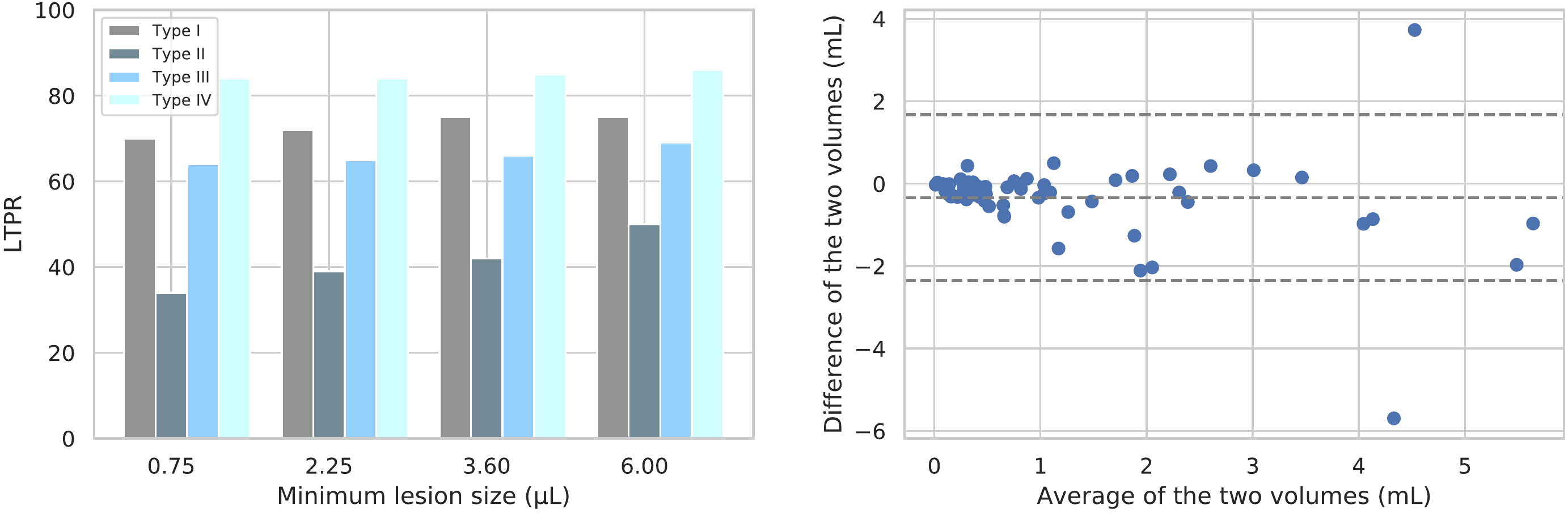}
\caption{On the left, the CL LTPR depending on the minimum lesion size considered, per lesion type. On the right, Bland-Altman plot (reference - prediction) of the manually and automatically segmented CL volumes.} \label{fig5}
\end{figure}

\textbf{Lesion-wise analysis.}
Evaluation metrics of U-Net\textsuperscript{-}, Multi-task U-Net\textsuperscript{-}, and Multi-task U-Net\textsuperscript{-} + ICD are reported in Table \ref{tab1}. The latter outperforms the others in all four metrics. This network achieves a CLs LTPR of 67\%, LFPR of 42\%, AVD of 36\%, and CLs classification accuracy of 86\%. We analyse the Multi-task U-Net\textsuperscript{-} + ICD LTPR depending on the minimum lesion size and per lesion type (see Figure \ref{fig5}). As can be observed, CLs Type II are the most challenging ones, with a detection rate of 34\% considering 0.75 \si{\micro\liter} as minimum lesion volume. Increasing the minimum lesion volume to 6\si{\micro\liter} (as in \cite{fartaria2019}), the network reaches an overall CL detection rate of 75\%, with consistent improvements especially for type II and type III CLs. A qualitative example of the segmentation outputs is shown in Figure \ref{fig6} in comparison with the experts' ground truth.
\begin{table} [hbt!]
\centering
    {\caption{Lesion-wise TPR and FPR, patient-wise AVD, and CL classification accuracy reported for the different networks. Statistical differences are found for the AVD between all three methods (p-values $<$ 0.01).}\label{tab1}} %
    {\begin{tabular}{l  c  c c c}
    \hline
    Network\hspace{5pt} &  \hspace{3pt}LTPR\hspace{3pt} & LFPR\hspace{3pt} &  AVD\hspace{3pt} & Accuracy \\
    \hline
    \hline
    3D U-Net\textsuperscript{-}  & 0.63 & 0.53 & 1.22 &   0.82 \\ \hline
    Multi-task 3D U-Net\textsuperscript{-}  & 0.66 & 0.44 & 0.55 & 0.85 \\ \hline
    Multi-task 3D U-Net\textsuperscript{-} + ICD  & \textbf{0.67} & \textbf{0.42} & \textbf{0.36} & \textbf{0.86} \\
    \hline
    \end{tabular}}
\end{table}

\textbf{False positives analysis.} Given the difficulty of the CL detection task, even for experts, we further analyzed the false positive lesions given by our best network architecture. Retrospectively, one of the experts re-evaluated each of the FP and assigned 24\% of those (385 lesions) to actual CLs or ``possible CLs"\fxnote{the possible CLs could be useful to train a dense label map, in which possible would have e.g. value 0.5}. This is on one side clear evidence of the difficulty of the task, as two experts missed them, and on the other side, a sign of the practical value of the automatic method proposed, for instance to present candidate lesions to support and speed up the experts' routine MRI analysis. 

\begin{figure} [hbt!]
\centering
\includegraphics[width=0.9999\textwidth]{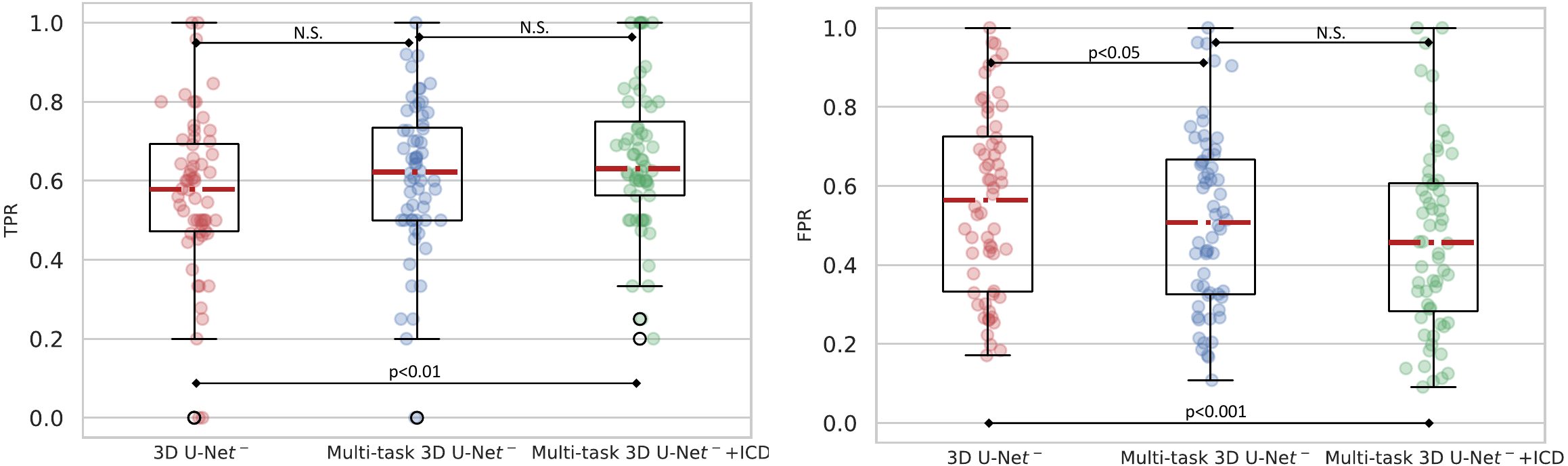}
\caption{TPR and FPR patient-wise for the three networks. N.S.: not significant.} \label{fig4}
\end{figure}

\textbf{Patient-wise analysis.}
The final architecture outperformed the two comparison networks also patient-wise (Figure \ref{fig4}). Adding the output tissue segmentation map helped reduce false positives in the WM, and therefore the overall FPR. Statistical differences were found between multi-task 3D U-Net\textsuperscript{-} + ICD and the baseline 3D U-Net\textsuperscript{-} for both LTPR and LFPR. Bland-Altman plot of the manually and automatically CLs volume segmented per patient is presented in Figure \ref{fig5} right. Aside from two outliers, all differences are within mean $\pm$ 1.96 SD, and we do not observe any systematic error estimation bias as a function of the total lesion volume size.

\section{Conclusion}
In this work, we explore the capability of deep learning-based techniques for the automated detection of MS cortical lesions in advanced MRI sequences (MP2RAGE, T2*w EPI, and T2*w GRE) at 7T. To the best of our knowledge, this is the first automated method specifically proposed for MS CL detection at ultra-high field MRI. Furthermore, it differentiates them into two classes: leukocortical and intracortical/subpial. 

Our work is evaluated on a large dataset of 60 MS patients including over 2000 CLs  manually labelled by two experts. Considering as baseline a simplified 3D U-Net, we increased the complexity of the classification task by simultaneously predicting GM and WM tissue types. Moreover, we propose an input channel dropout technique to tackle the issues of artifacts or missing parts of the brain in the T2*w contrasts. This architecture achieves a 67\% CL detection rate with 42\% false positives, considering a minimum lesion size of 0.75 \si{\micro\liter}, much lower than any previous work. Interestingly, a retrospective analysis of the false positives by a single expert showed that 24\% of them could be considered CLs or possible CLs missed in the initial labelling. This proves the potential of the proposed method for supporting experts in the tedious process of CL labelling, possibly presenting them with candidate lesions.  

Future work will include exploring the T1 quantitative map of the MP2RAGE sequence, and experimenting during training a soft ground truth, which could help coping with the experts' definition of ``possible cortical lesion".

\subsubsection{Acknowledgments.}
This project is supported by the European Union's Horizon 2020 research and innovation program under the Marie Sklodowska-Curie project TRABIT (agreement No 765148), the Centre d$'$Imagerie BioM\'{e}dicale of the University of Lausanne, the Swiss Federal Institute of Technology Lausanne, the University of Geneva, the Centre Hospitalier Universitaire Vaudois, and the H\^{o}pitaux Universitaires de Gen\`{e}ve. Erin S Beck is supported by a Career Transition Fellowship from the National Multiple Sclerosis Society. Pascal Sati, Erin S Beck, and Daniel S Reich are supported by the Intramural Research Program of the National Institute of Neurological Disorders and Stroke, National Institutes of Health, Bethesda, Maryland, USA. Ahmed Abdulkadir is supported by the Swiss National Science Foundation grant SNSF 173880. 


\bibliographystyle{splncs04}
\bibliography{paper655}

\begin{thebibliography}{10}
\providecommand{\url}[1]{\texttt{#1}}
\providecommand{\urlprefix}{URL }
\providecommand{\doi}[1]{https://doi.org/#1}

\bibitem{ants}
Avants, B.B., Tustison, N., Song, G.: Advanced normalization tools (ants).
  Insight j  \textbf{2}(365),  1--35 (2009)

\bibitem{erin}
Beck, E.S., Sati, P., Sethi, V., Kober, T., Dewey, B., Bhargava, P., Nair, G.,
  Cortese, I.C., Reich, D.S.: Improved visualization of cortical lesions in
  multiple sclerosis using 7t mp2rage. American Journal of Neuroradiology
  \textbf{39}(3),  459--466 (2018)

\bibitem{calabrese}
Calabrese, M., Filippi, M., Gallo, P.: Cortical lesions in multiple sclerosis.
  Nature Reviews Neurology  \textbf{6}(8), ~438 (2010)

\bibitem{challenge}
Carass, A., Roy, S., Jog, A., Cuzzocreo, J.L., Magrath, E., Gherman, A.,
  Button, J., Nguyen, J., Prados, F., Sudre, C.H., et~al.: Longitudinal
  multiple sclerosis lesion segmentation: resource and challenge. NeuroImage
  \textbf{148},  77--102 (2017)

\bibitem{3dunet}
{\c{C}}i{\c{c}}ek, {\"O}., Abdulkadir, A., Lienkamp, S.S., Brox, T.,
  Ronneberger, O.: 3d u-net: learning dense volumetric segmentation from sparse
  annotation. In: International conference on medical image computing and
  computer-assisted intervention. pp. 424--432. Springer (2016)

\bibitem{reuben}
Dorent, R., Li, W., Ekanayake, J., Ourselin, S., Vercauteren, T.: Learning
  joint lesion and tissue segmentation from task-specific hetero-modal
  datasets. arXiv preprint arXiv:1907.03327  (2019)

\bibitem{mj}
Fartaria, M.J., Bonnier, G., Roche, A., Kober, T., Meuli, R., Rotzinger, D.,
  Frackowiak, R., Schluep, M., Du~Pasquier, R., Thiran, J.P., et~al.: Automated
  detection of white matter and cortical lesions in early stages of multiple
  sclerosis. Journal of Magnetic Resonance Imaging  \textbf{43}(6),  1445--1454
  (2016)

\bibitem{mj_pv}
Fartaria, M.J., Roche, A., Meuli, R., Granziera, C., Kober, T., Cuadra, M.B.:
  Segmentation of cortical and subcortical multiple sclerosis lesions based on
  constrained partial volume modeling. In: International Conference on Medical
  Image Computing and Computer-Assisted Intervention. pp. 142--149. Springer
  (2017)

\bibitem{fartaria2019}
Fartaria, M.J., Sati, P., Todea, A., Radue, E.W., Rahmanzadeh, R., O'Brien, K.,
  Reich, D.S., Cuadra, M.B., Kober, T., Granziera, C.: Automated detection and
  segmentation of multiple sclerosis lesions using ultra--high-field mp2rage.
  Investigative radiology  \textbf{54}(6),  356--364 (2019)

\bibitem{review}
Garc{\'\i}a-Lorenzo, D., Francis, S., Narayanan, S., Arnold, D.L., Collins,
  D.L.: Review of automatic segmentation methods of multiple sclerosis white
  matter lesions on conventional magnetic resonance imaging. Medical image
  analysis  \textbf{17}(1),  1--18 (2013)

\bibitem{niftynet}
Gibson, E., Li, W., Sudre, C., Fidon, L., Shakir, D.I., Wang, G., Eaton-Rosen,
  Z., Gray, R., Doel, T., Hu, Y., et~al.: Niftynet: a deep-learning platform
  for medical imaging. Computer methods and programs in biomedicine
  \textbf{158},  113--122 (2018)

\bibitem{kaur}
Kaur, A., Kaur, L., Singh, A.: State-of-the-art segmentation techniques and
  future directions for multiple sclerosis brain lesions. Archives of
  Computational Methods in Engineering pp. 1--27 (2020)

\bibitem{kilsdonk}
Kilsdonk, I.D., Jonkman, L.E., Klaver, R., van Veluw, S.J., Zwanenburg, J.J.,
  Kuijer, J.P., Pouwels, P.J., Twisk, J.W., Wattjes, M.P., Luijten, P.R.,
  et~al.: Increased cortical grey matter lesion detection in multiple sclerosis
  with 7 t mri: a post-mortem verification study. Brain  \textbf{139}(5),
  1472--1481 (2016)

\bibitem{kober}
Kober, T., Granziera, C., Ribes, D., Browaeys, P., Schluep, M., Meuli, R.,
  Frackowiak, R., Gruetter, R., Krueger, G.: Mp2rage multiple sclerosis
  magnetic resonance imaging at 3 t. Investigative radiology  \textbf{47}(6),
  346--352 (2012)

\bibitem{larosa2020}
La~Rosa, F., Abdulkadir, A., Fartaria, M.J., Rahmanzadeh, R., Lu, P.J.,
  Galbusera, R., Barakovic, M., Thiran, J.P., Granziera, C., Bach~Cuadra, M.:
  Multiple sclerosis cortical and {WM} lesion segmentation at {3T} {MRI}: a
  deep learning method based on {FLAIR} and {MP2RAGE}. NeuroImage: Clinical p.
  102335 (Jun 2020). \doi{10.1016/j.nicl.2020.102335},
  \url{https://linkinghub.elsevier.com/retrieve/pii/S2213158220301728}

\bibitem{larosa2019}
La~Rosa, F., Fartaria, M.J., Abdulkadir, A., Rahmanzadeh, R., Lu, P.J.,
  Galbusera, R., Granziera, C., Thiran, J.P., Bach~Cuadra, M.: Deep
  learning-based detection of cortical lesions in multiple sclerosis patients
  with flair, dir, and mp2rage mri sequences. Multiple Sclerosis Journal
  \textbf{25}(CONF),  131--356 (2019)

\bibitem{magliozzi}
Magliozzi, R., Howell, O.W., Reeves, C., Roncaroli, F., Nicholas, R., Serafini,
  B., Aloisi, F., Reynolds, R.: A gradient of neuronal loss and meningeal
  inflammation in multiple sclerosis. Annals of neurology  \textbf{68}(4),
  477--493 (2010)

\bibitem{maranzano}
Maranzano, J., Dadar, M., Rudko, D., De~Nigris, D., Elliott, C., Gati, J.,
  Morrow, S., Menon, R., Collins, D., Arnold, D., et~al.: Comparison of
  multiple sclerosis cortical lesion types detected by multicontrast 3t and 7t
  mri. American Journal of Neuroradiology  \textbf{40}(7),  1162--1169 (2019)

\bibitem{mp2rage}
Marques, J.P., Kober, T., Krueger, G., van~der Zwaag, W., Van~de Moortele,
  P.F., Gruetter, R.: Mp2rage, a self bias-field corrected sequence for
  improved segmentation and t1-mapping at high field. Neuroimage
  \textbf{49}(2),  1271--1281 (2010)

\bibitem{sati}
Sati, P., Thomasson, D., Li, N., Pham, D., Biassou, N., Reich, D., Butman, J.:
  Rapid, high-resolution, whole-brain, susceptibility-based mri of multiple
  sclerosis. Multiple Sclerosis Journal  \textbf{20}(11),  1464--1470 (2014)

\bibitem{sati2}
Sati, P., Patil, S., Inati, S., Wang, W.T., Derbyshire, J.A., Krueger, G.,
  Reich, D.S., Butman, J.A.: Rapid mr susceptibility imaging of the brain using
  segmented 3d echo-planar imaging (3d epi) and its clinical applications.
  Magnetom FLASH  \textbf{68},  26--32 (2017)

\bibitem{criteria2017}
Thompson, A.J., Banwell, B.L., Barkhof, F., Carroll, W.M., Coetzee, T., Comi,
  G., Correale, J., Fazekas, F., Filippi, M., Freedman, M.S., et~al.: Diagnosis
  of multiple sclerosis: 2017 revisions of the mcdonald criteria. The Lancet
  Neurology  \textbf{17}(2),  162--173 (2018)

\bibitem{treaba}
Treaba, C.A., Granberg, T.E., Sormani, M.P., Herranz, E., Ouellette, R.A.,
  Louapre, C., Sloane, J.A., Kinkel, R.P., Mainero, C.: Longitudinal
  characterization of cortical lesion development and evolution in multiple
  sclerosis with 7.0-t mri. Radiology  \textbf{291}(3),  740--749 (2019)

\bibitem{n4}
Tustison, N.J., Avants, B.B., Cook, P.A., Zheng, Y., Egan, A., Yushkevich,
  P.A., Gee, J.C.: N4itk: improved n3 bias correction. IEEE transactions on
  medical imaging  \textbf{29}(6),  1310--1320 (2010)

\end{thebibliography}

\end{document}